\documentclass[12pt]{article}

\usepackage{amsmath,amsfonts,amssymb,color,graphicx,overpic,}
\usepackage{slashed}
\usepackage{epsfig}
\usepackage[utf8]{inputenc}
\usepackage{amsmath}
\usepackage{ wasysym }
\usepackage{graphicx}
\usepackage[english]{babel}
\usepackage{graphic x}
\usepackage{relsize}
\usepackage{cite}

\begin{document}

\begin{titlepage}
\rightline{July 2018}
\vskip 1.9cm
\centerline{\large \bf
Direct detection of mirror helium dark matter
}
\vskip 0.3cm
\centerline{\large \bf
in the CRESST-III experiment}

\vskip 2.4cm
\centerline{R. Foot\footnote{
E-mail address: rfoot@unimelb.edu.au}}

\vskip 0.4cm
\centerline{\it ARC Centre of Excellence for Particle Physics at the Terascale,}
\centerline{\it School of Physics, The University of Sydney, NSW 2006, Australia}
\vskip 0.1cm
\centerline{and}
\vskip 0.1cm
\centerline{\it ARC Centre of Excellence for Particle Physics at the Terascale,}
\centerline{\it School of Physics, The University of Melbourne, VIC 3010, Australia
}

\vskip 2.5cm
\noindent
Within the context of mirror dark matter, the
dominant mass component of the Milky Way dark halo consists of mirror helium ions.
Mirror helium  can interact
with ordinary matter if the kinetic mixing interaction exists.
Mirror helium being rather light, $m \simeq 3.73$ GeV, generally produces sub-keV recoils
in direct detection experiments. Recently, the CRESST-III experiment  
has began probing the sub-keV recoil energy region
and is currently the most sensitive probe of such particles.
We point out here that the small excess seen in the low energy recoil data obtained
in the CRESST-III experiment 
is consistent with
mirror helium scattering if the kinetic mixing parameter is
around $\epsilon \approx 5 \times 10^{-10}$.
This kinetic mixing strength lies within the estimated range favoured by small scale structure considerations.

\end{titlepage}

The mirror dark matter model supposes that dark matter arises from a hidden sector which is an exact copy of
the standard model. That is, the
Lagrangian describing fundamental physics is
\begin{eqnarray}
{\cal L} = {\cal L}_{\rm SM}(e,u,d,\gamma,...) + {\cal L}_{\rm SM}(e',u',d',\gamma',...)
+ {\cal L}_{\rm mix}
\ .
\label{yyy6}
\end{eqnarray}
The Lagrangian features an exact unbroken $Z_2$ symmetry, which can be interpreted as space-time parity if the
chirality of the hidden sector  is flipped \cite{flv}.
The mirror sector particles interact with the standard ones via gravity and via the
kinetic mixing interaction \cite{holdom,fhe}, which also leads to photon - mirror photon kinetic mixing:
\begin{eqnarray}
{\cal L}_{\rm mix} = \frac{\epsilon}{2}  F^{\mu \nu} F^{'}_{\mu \nu}
\ .
\label{mix1}
\end{eqnarray}
Here $F^{\mu \nu}$ [$F^{'}_{\mu \nu}$] is the field strength tensor of the photon
[mirror photon].
The kinetic mixing induces tiny ordinary electric charges for the charged
mirror sector particles, of $\pm \epsilon e$ for the mirror proton and mirror electron respectively.

In this picture, the mirror matter particles, the mirror nuclei and mirror electrons, constitute the inferred dark matter in the Universe.
It has been argued in a number of papers that such dark matter is consistent with both large and small scale structure
observations provided that the kinetic mixing strength is in the range:
$\epsilon = (1-6)\times 10^{-10}$.
The reader is referred to \cite{sph} and recent papers \cite{sunny2,foot18a,foot18} (for small scale structure), \cite{Ber,IV,foot07} (for large scale structure); see also   
the review \cite{review} for a more detailed bibliography.

The halo of the Milky Way consists of an ionized plasma
containing mirror electrons ($e'$), mirror hydrogen ions (H$'$), mirror helium ions (He$'$), and potentially also a mirror 
metal component.
In principle, all these components can be searched for in direct detection experiments \cite{footold,footr,jc2}. 
In practice the problem is potentially quite complex due to shielding effects of Earth-bound dark matter \cite{jc1}.
This is especially true for mirror electron scattering
as mirror electrons are so light and easily influenced by the induced mirror electromagnetic fields.
Still,
 mirror electron
scattering off atomic electrons has been advocated \cite{jc1,fr} as a viable explanation of the DAMA annual modulation signal \cite{dama1,dama2,dama3}, 
consistent with the kinetic mixing
value favoured by small scale structure considerations.

Previous work has paid relatively little attention to the light ion components, mirror hydrogen and mirror helium,
as these components produce sub-keV nuclear recoils, typically below the threshold of current direct detection experiments.
However, these components are rather important, with mirror helium
of mass $m \simeq 3.73$ GeV  
expected to be the dominant mass component of the Milky Way halo \cite{Ber,fp,review}.  \footnote{Units with $\hbar = c = 1$ are used unless otherwise specified.}
Recently, the CRESST-III experiment  \cite{cresstIII}
has began  exploring the sub-keV recoil energy region
and is currently the most sensitive probe of the halo mirror helium component.
The CRESST-III collaboration has measured a rising event rate in the nuclear recoil band at low recoil energies down to a threshold of 0.1 keV. 
We point out here that this excess 
can be explained by mirror helium scattering if the kinetic mixing parameter is
around $\epsilon \approx 5 \times 10^{-10}$, 
which lies within the estimated range
favoured by small scale structure considerations.

The small kinetic mixing induces tiny electric charges for the mirror particles. This means that a mirror helium (He$'$) ion can Rutherford scatter
off an ordinary nucleus $A$ of atomic number $Z$. Considering a He$'$ ion of velocity $v$, scattering off a nucleus at rest, then the cross section is
\begin{eqnarray}
\frac{d\sigma}{dE_R} = \frac{\lambda_{\rm He'}  
F_A^2 F_{\rm He'}^2}
{E_R^2 v^2}
\label{cs}
\end{eqnarray}
where $\lambda_{\rm He'}  \equiv 8\pi \epsilon^2 Z^2\alpha^2/m_A$,
and $E_R$ is the recoil energy of the scattered nucleus.
Here, $F_A$, $F_{\rm He'}$ are the form factors that account for the finite size of the ordinary and mirror nuclei.
In the numerical work we use the Helm parameterization of the form factors \cite{helm}.

The mirror helium interaction rate in a direct detection experiment takes the form:
\begin{eqnarray}
\frac{dR}{dE_R} &=&
N_T  n_{\rm He'}
\int \frac{d\sigma}{dE_R}
\ f({\textbf{v}}; {\textbf{v}}_E)\ |{\textbf{v}}| \
d^3v \nonumber \\
&=& N_T  n_{\rm He'}
\frac{\lambda_{\rm He'}}{E_R^2} 
F_A^2 F_{\rm He'}^2
\ I({\textbf{v}}_E)
\label{55}
\end{eqnarray}
where
\begin{eqnarray}
I({\textbf{v}}_E)
\equiv \int^{\infty}_{|{\bf{v}}| > v_{min} (E_R)}
\ \frac{f({\textbf{v}}; {\textbf{v}}_E)}{|{\textbf{v}}|} \ d^3 v
\ .
\label{III}
\end{eqnarray}
Here, $N_T$ is the number of target atoms per kg of detector,
$v_{\rm min} \ = \ \sqrt{E_R m_A/2\mu^2}$, where $\mu$ is the $A$ - He$'$ reduced mass,
and $n_{\rm He'}$ is the halo mirror helium  number density at the detector's location.
Also, $f({\textbf{v}}; {\textbf{v}}_E)$
is the velocity distribution of the halo mirror helium ions which arrive at the detector.
As indicated, this distribution will depend on the velocity of the halo wind as measured from Earth
[${\textbf{v}}_E (t) $]. 
If shielding effects are important, then this distribution can also
depend on 
the angle between the direction of the halo wind
and the zenith at the detector's location
and possibly also depend on the detector's geographical location.

We assume shielding effects are not of critical importance for the average rate of mirror helium interactions, and take the halo mirror helium distribution to be a boosted
Maxwellian:
\begin{eqnarray}
f({\textbf{v}};{\textbf{v}}_E) = \left( \frac{1}{\pi v_0^2}\right)^{\frac{3}{2}}
\ exp \left( \frac{ -({\textbf{v}}+{\textbf{v}}_E)^2}
{v_0^2}\right)
\ .
\end{eqnarray}
The quantity $v_0$ is
\begin{eqnarray}
v_0 = v_{\rm rot}\sqrt{\frac{\bar m}{m_{\rm He'}}}
\end{eqnarray}
where $v_{\rm rot} \approx 220$ km/s is the asymptotic halo rotational velocity of the Milky Way, and $\bar m \approx 1.1$ GeV is the mean particle mass of the halo, e.g. \cite{review}.
For the purposes of this paper, we consider the average scattering
rate with $v_E = v_{rot} + 12 {\rm km/s}$, and take
$m_{\rm He} n_{\rm He'} = 0.3$ GeV/cm$^3$.

\begin{figure}[t]
  \begin{minipage}[b]{0.5\linewidth}
    \centering
    \includegraphics[width=0.7\linewidth,angle=270]{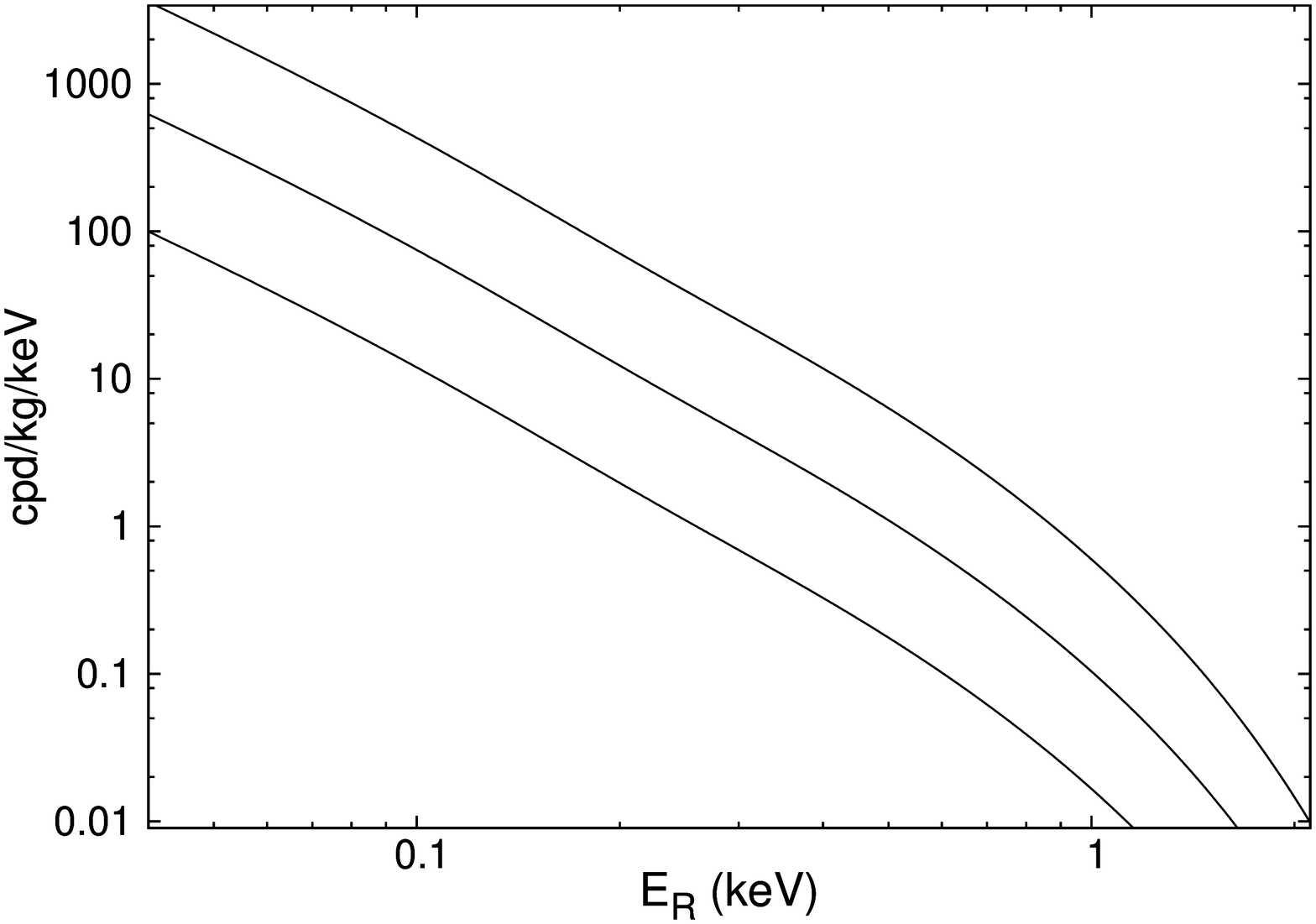}
     (a)
    \vspace{4ex}
  \end{minipage}
  \begin{minipage}[b]{0.5\linewidth}
    \centering
    \includegraphics[width=0.7\linewidth,angle=270]{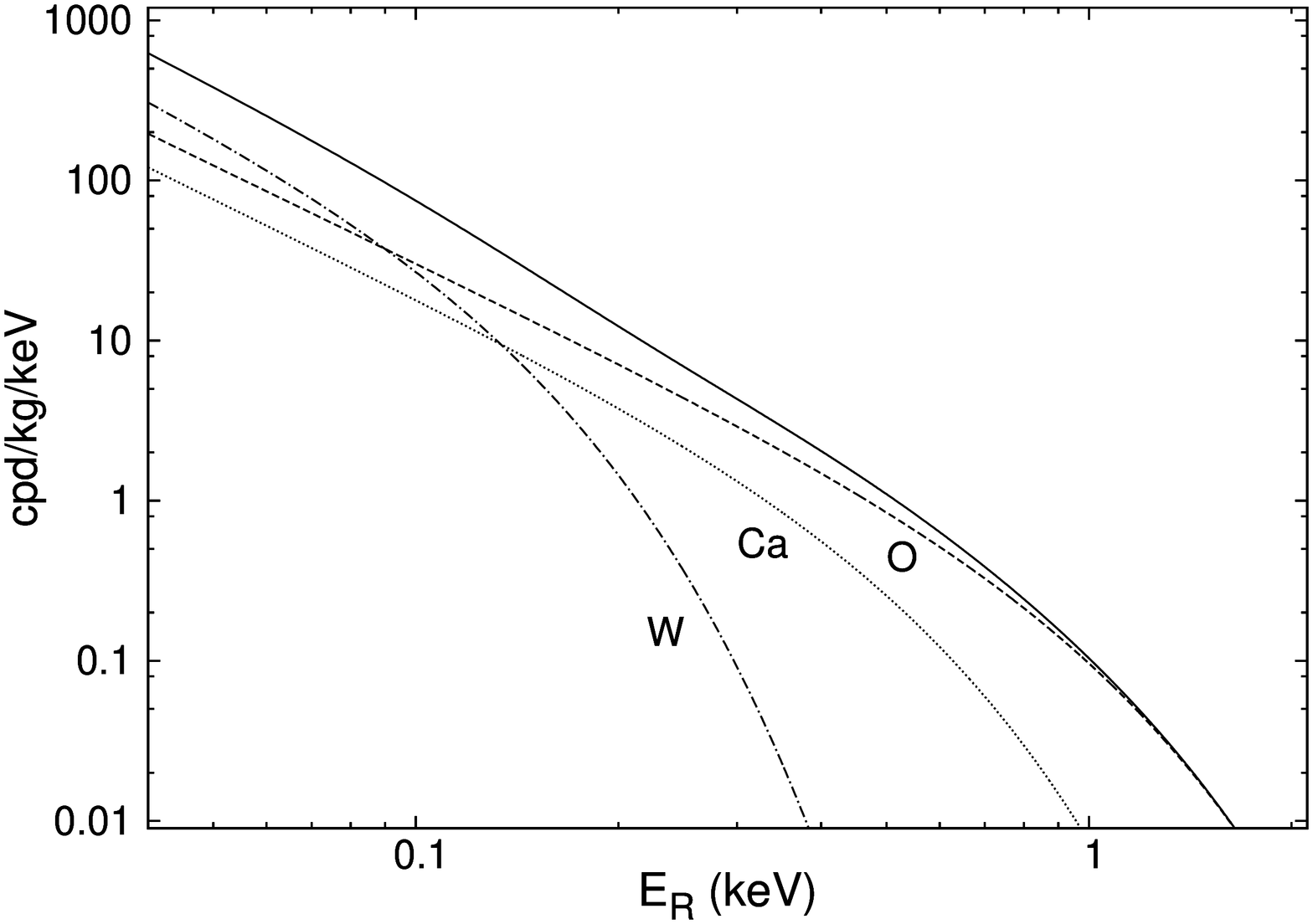}
    (b)
    \vspace{4ex}
  \end{minipage}
\vskip -1.1cm
\caption{
\small
(a) Scattering rate of halo mirror helium ions on a CaWO$_4$ target.
The bottom, middle and top lines assume kinetic mixing of $\epsilon/10^{-10} = 1, \ 2.5, \ 6$.
(b) Decomposition of the scattering rate in terms of the oxygen (dashed), calcium (dotted) and tungsten (dashed-dotted) components
for $\epsilon = 2.5\times 10^{-10}$.
}
\end{figure}

In Figure 1 we give the halo mirror helium scattering rate for a CaWO$_4$ target crystal. 
For this figure,
an idealized detector with perfect energy resolution and detection efficiency is assumed.
The normalization of the rate is proportional to $\epsilon^2$ and we consider three
kinetic mixing values: $\epsilon/10^{-10} = 1, \ 2.5,\ 6$, 
broadly covering the possible $\epsilon$   range of interest identified from small scale structure considerations.
In Figure 1b the decomposition of the scattering rate into the oxygen, calcium and tungsten target components is shown for $\epsilon = 2.5\times 10^{-10}$.

In Figure 2 we evaluate the halo mirror helium scattering rate  per 0.025 keV for the CRESST-III phase 1 exposure of $2.39$ 
kg days. For Figure 2, a  detection efficiency of $0.75\times 0.5$ is assumed for recoils off oxygen, while an efficiency of $0.75$ is assumed
for recoils off calcium and tungsten. These values roughly match those of the
CRESST-III nuclear recoil acceptance region.  In Figure 2 we have taken the energy resolution in the low energy region to be $\sigma/E_R = 0.1$, for definiteness.
A kinetic mixing value of $\epsilon = 5.2\times 10^{-10}$ was found
to roughly match the CRESST-III phase 1 data in the nuclear recoil band.

\begin{figure}[t]
    \centering
    \includegraphics[width=0.46\linewidth,angle=270]{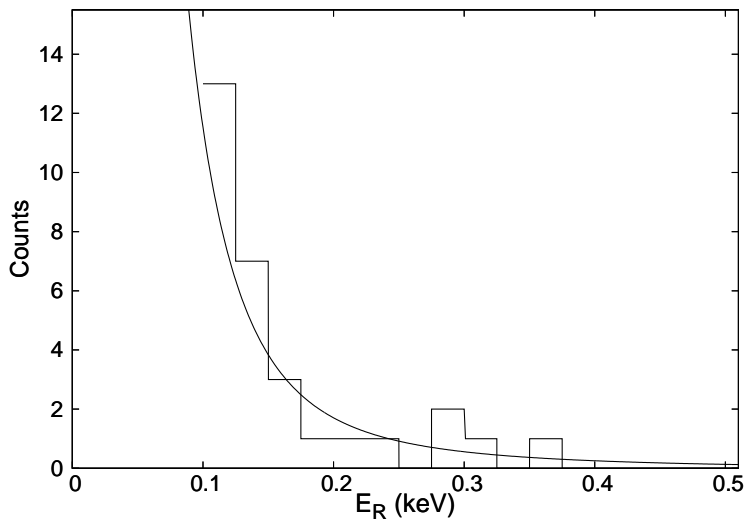}
     \vspace{-1ex}
    \centering
\caption{
\small
Scattering rate of halo mirror helium ions on a CaWO$_4$ crystal per 0.025 keV for the CRESST-III phase 1 exposure of $2.39$ 
kg days for $\epsilon = 5.2 \times 10^{-10}$. Also shown are the CRESST-III phase 1 data. 
}
\end{figure}

Figure 2 indicates that the low energy excess in the CRESST-III phase 1 data is compatible with halo mirror helium interactions.
Naturally, there are a number of significant uncertainties (e.g. number density estimate of He$'$ at the detector, experimental issues such as uncertain
detector energy calibration and resolution etc.) and assumptions (e.g. collisional shielding can be neglected).
Considering the latter, collisional shielding of the detector due to Earth-bound He$'$ dark matter can potentially suppress
the halo He$'$ interaction rate in the detector. The importance of collisional shielding can be estimated from the density of Earth-bound 
He$'$ at the Earth's surface, $n_{\rm He'}(R_E)$. Following a calculation similar to that done in \cite{fr} we find that  collisional shielding
will not significantly suppress the average rate provided that $n_{\rm He'}(R_E) \lesssim few \times 10^{10}\ {\rm cm^{-3}}$. Of course, values of 
$n_{\rm He'}(R_E) \gtrsim 10^{10}\ {\rm cm^{-3}}$
can still lead to important effects, including large diurnal modulation for 
nuclear recoils.\footnote{If $n_{\rm He'}(R_E) \lesssim few \times 10^{10}\ {\rm cm^{-3}}$ then
collisional shielding will be relatively unimportant also for 
halo mirror electron scattering off target electrons. 
If the DAMA experiment is to be explained by mirror electron induced electron recoils then shielding effects appear to be
necessary \cite{fr}, but might be due (mainly) to induced mirror electromagnetic fields.}

To conclude, in the mirror dark matter model, the dominant halo mass component of the Milky Way halo is mirror helium, a particle of mass $3.73$ GeV.
Such particles can interact with ordinary matter via the kinetic mixing interaction, expected to exist at the level of $\epsilon \approx {\rm few} \times 10^{-10}$
for consistent small scale structure.
Mirror helium is so light that only sub-keV nuclear recoils are expected, i.e. interactions below the threshold of most current direct detection experiments.
However, the  CRESST-III collaboration has began probing the sub-keV recoil 
region, and has measured a rising event rate in the nuclear recoil band at low energies down to a 
threshold of 0.1 keV. This excess has been interpreted here as halo mirror helium interactions, with
a kinetic mixing strength of around $\epsilon \approx 5 \times 10^{-10}$ implicated. 
This kinetic mixing strength is roughly consistent with the estimates obtained from small scale structure considerations.
While the low energy excess apparent in the CRESST-III data, appears to be compatible with expectations from halo mirror helium interactions, it
will be important to measure an annual modulation, and potentially also a diurnal modulation, to firmly establish that the excess is due
to dark matter interactions.

\vskip 0.4cm
\noindent
{\large \bf Acknowledgments}

\vskip 0.2cm
\noindent
This work was supported by the Australian Research Council.


\end{document}